\documentclass{osa-article}

\journal{oe}

\articletype{Research Article}

\begin{document}

\title{Lithography-free Kirchhoff's Metasurfaces}

\author{Takuhiro Kumagai,\authormark{1} Naoki To,\authormark{1}, Armandas Bal\v{c}ytis,\authormark{1,2} Gediminas Seniutinas\authormark{3}, Saulius Juodkazis\authormark{3,4,5}\\\protect Yoshiaki Nishijima\authormark{1,4}}

\address{\authormark{1}Department of Physics, Electrical and Computer Engineering, Graduate School of Engineering, Yokohama National University, 79-5 Tokiwadai, Hodogaya-ku, Yokohama 240-8501, Japan\\
\authormark{2}Center for Physical Sciences and Technology, A. Go\v{s}tauto 9, LT-01108 Vilnius, Lithuania\\
\authormark{3}Optical Sciences Centre and ARC Training Centre in Surface Engineering for Advanced Materials (SEAM), School of Science, Swinburne University of Technology, Hawthorn, VIC 3122, Australia\\
\authormark{4}Institute of Advanced Sciences, Yokohama National University, 79-5 Tokiwadai, Hodogaya-ku, Yokohama 240-8501, Japan.\\
\authormark{5}Tokyo Tech World Research Hub Initiative (WRHI), School of Materials and Chemical Technology, Tokyo Institute of Technology, 2-12-1, Ookayama, Meguro-ku, Tokyo 152-8550, Japan
}

\email{\authormark{*}nishijima-yoshiaki-sp@ynu.ac.jp} 



\begin{abstract}
Lithography-free metasurfaces composed of a nano-layered stack of materials are attractive not only due to their optical properties but also by virtue of fabrication simplicity and cost reduction of devices based on such structures. We demonstrate a multi-layer metasurface with engineered electromagnetic absorption in the mid-infrared (MIR) wavelength range. Characterization of thin SiO$_2$ and Si films sandwiched between two Au layers by way of experimental absorption and thermal radiation measurements as well as finite difference time domain (FDTD) numerical simulations is presented. Comparison of experimental and simulation data of optical properties of multilayer metasurfaces show guidelines for the absorber/emitter applications. 
\end{abstract}

\section{Introduction}

Metasurfaces, composed of patterns of sub-wavelength antennas, are used for tailoring spectral, polarisation and angular dependencies of electromagnetic radiation absorption as well as thermal radiation emission. However, nano-structure arrays typical for such applications necessitate the use of high resolution lithography systems, e.g., electron beam lithography (EBL), size reduction photo-lithography based on steppers or focused ion beam (FIB) nanofabrication~\cite{Azad}. Therefore, a lithography-free approach for fabrication of metasurfaces has inherent advantages, especially as it eschews the considerable up-front cost of high resolution patterning equipment, that gets incorporated into the end-user price of the fabricated devices, in particular, when macroscopic area coverage has to be achieved~\cite{Hedayati}.

There are several lithography-free approaches for fabricating structures with metasurface properties. One of the most notable is based on the Tamm plasmon polaritons supported by a thin metal film on a multi-layer Bragg reflector stack~\cite{Yang,Yang1,Wang,Wang1,Liu}. So called plasmon super absorbers, on the other hand, are comprised of a metal ground plate and a thin overlaying semiconductor film~\cite{Dias,Park1}. For optimisation of the design of metasurfaces for reduced reflectivity and transmittance, the optical impedance matching method can be used~\cite{Michael}, which is particularly promising when the refractive index $\sqrt{\varepsilon}\equiv\tilde{n} = n + i\kappa$ of the actual thin film is known, here $\varepsilon$ is the permittivity.  
\begin{figure}[tb]
\begin{center}
\includegraphics[width=10cm]{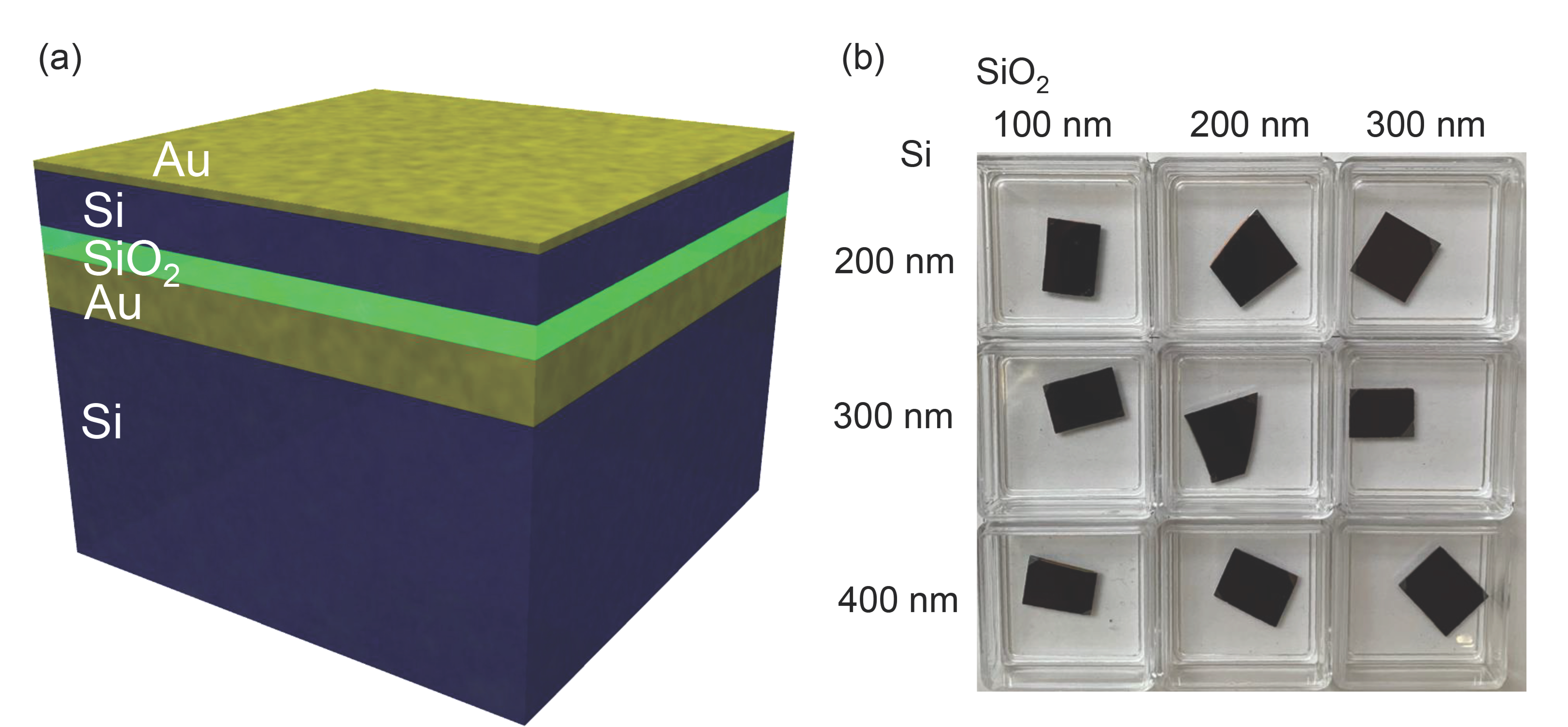}
\caption{(a) Schematic illustration of fabricated multi-layer metasurfaces. (b) Photographs of the absorber samples with different semiconductor and dielectric layers.} \label{f1}
\end{center}
\end{figure}

Some target applications for plasmon absorbers include the detection of MIR radiation and as thermal emitters in the field of gas sensing~\cite{arxiv}. For this purpose, it is important to tune the spectral properties of absorption at specific wavelengths, which in conventional metamaterials is usually done by adjusting the geometry or placement of nanoantennas. Reciprocity between emittance, $E$, and absorbance, $A$, has been confirmed experimentally at the MIR spectral range~\cite{arxiv}. 

Here we experimentally demonstrate a MIR metamaterial composed of two different dielectric layers wedged between metal films, which is capable of electromagnetic absorption and emission in the wavelength region spanning from 2 to 10~$\mu$m. These lithography-free multi-layer structures simplify fabrication of metasurfaces that can be used for both spectrally selective thermal radiation emittance and infrared absorbance.

\section{Results and discussion}

\begin{figure}[tb]
\begin{center}
\includegraphics[width=15cm]{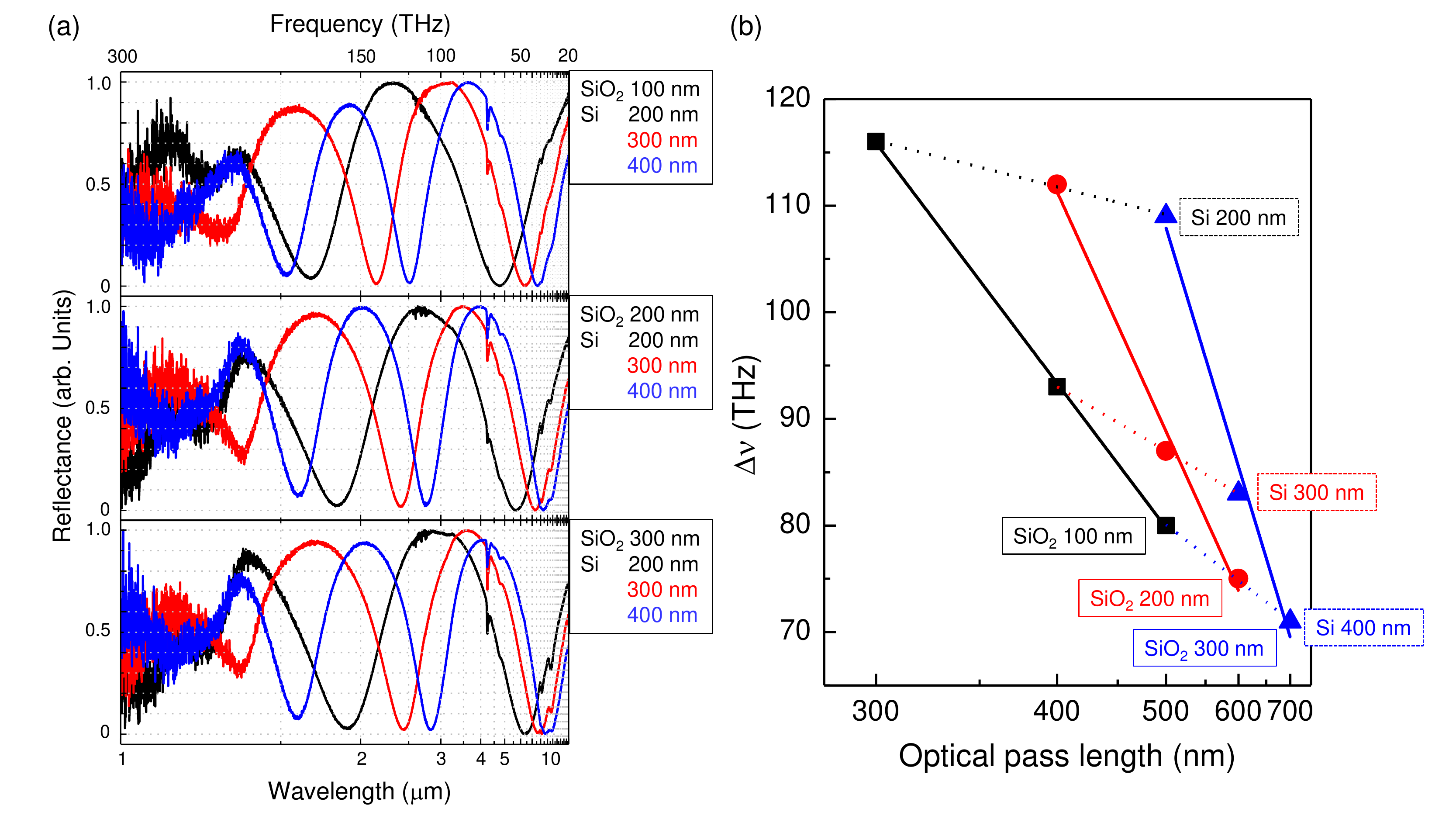}
\caption{(a) Reflectance spectra of the fabricated absorber samples with SiO$_2$ layer thickness changing from 100 to 300~nm and Si from 200 to 400~nm. (b) The full width at half maximum (FWHM) $\Delta\nu$ vs. optical path length, calculated as thickness of Si and SiO$_2$ layers at normal incidence. The slope of the dependence defines the group index $n_{g}$.} \label{f2}
\end{center}
\end{figure}
\begin{figure}[tb]
\begin{center}
\includegraphics[width=9cm]{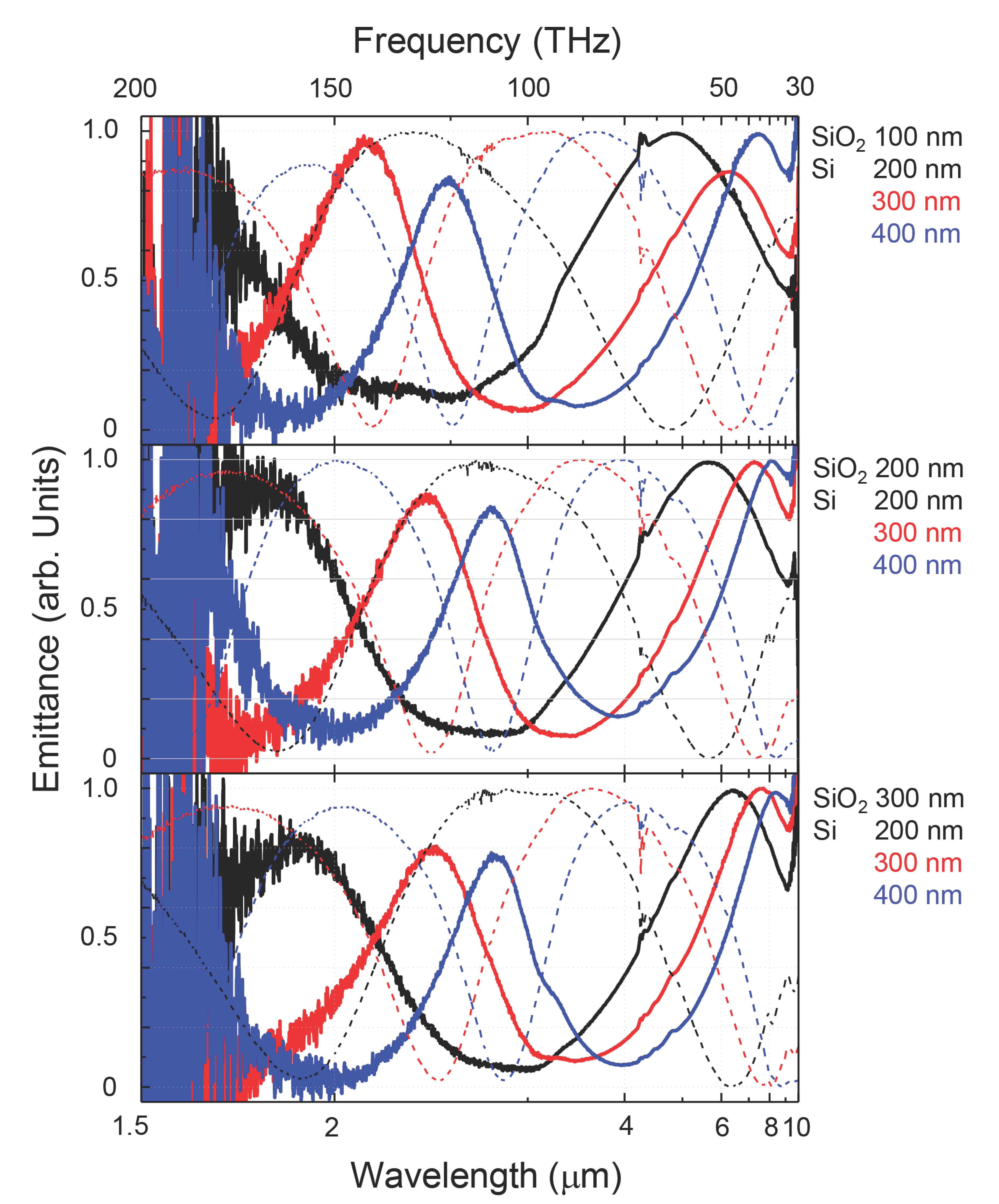}
\caption{Thermal radiation emission spectra of multilayer metasurfaces; dashed lines are the reflectance spectra from Fig.~\ref{f2}(a). Thermodynamic equivalence based on energy conservation for absorbance, reflectance, and transmittance $A + R + T = 1$ requires that emittance $E = 1-R = A$ in the absence of transmittance, as is in the case considered here due to a thick Au back reflector layer; Fig.~\ref{f1}(a)).  } \label{f3}
\end{center}
\end{figure}

\subsection{Experimental results}

A schematic sketch of the multi-layered metasurfaces with Si and SiO$_2$ layers is depicted in Fig.~\ref{f1}. Due to the strong adhesion between Si and SiO$_2$ films no additional intermediary layers were required. Conversely, 3~nm thickness Ti adhesion layers were deposited to form robust interfaces between Si and Au as well as SiO$_2$ and Au. Alternatively, chromium can also be used as an adhesive between Au and dielectrics, however, it is unsuitable for thermal emitter applications, as heating can result in Cr readily diffusing into Au as well as layer delamination. Moreover, spectral properties of emitters can be altered by the presence of even a thin layer of Cr. The material evaporation step was carried out over the surface of an 8-inch wafer and yielded a homogeneous metasurface coating, however, for simplified handling during experimental characterization, the Si substrate was cut into smaller pieces. Fabricated metasurfaces had a dark brown appearance, as shown in Fig.~\ref{f1}(b). All samples possessed two or three alternating absorption bands located in the visible wavelength region, and these higher order spectral signatures were the origin of the color.

Figure~\ref{f2}(a) shows the reflectance spectra of fabricated metasurfaces over a spectral range spanning from near-to-mid IR (1-10~$\mu$m wavelengths). Major absorbance peaks appeared in the range from 5 to 10~$\mu$m. Towards the longer MIR wavelengths reflection peaks become broader with full-width at half maximum (FWHM) of $\Delta\lambda\sim 5~\mu$m. In the frequency scale, most of the signatures possess comparable widths and uniform peak separations $\Delta\nu$, which are defined by the free spectral range (FSR): $FSR = c/n_g L$, where $c$ is the speed of light, $n_g$ is the group index, and $L$ is the optical path length. From analysis of FSR (Fig.~\ref{f2}(b)), $n_g$ was in the range from 5 to 9. Thinner SiO$_2$ and Si films result in a larger FSR, e.g.,  SiO$_2$ of 100~nm and  Si of 200 nm yields in $n_g$ = 8.6, whereas 300~nm SiO$_2$ and 400~nm  Si layer thicknesses correspond to $n_g$ = 5.5. Both SiO$_2$ and Si thickness were inversely proportional to the total optical path length (Fig.~\ref{f2}(b)). The influence of Si thickness on $\Delta\nu$ is much more pronounced than that of SiO$_2$ due to the higher refractive index. Experimentally measured absorption at the reflection dips was large and approached nearly the perfect absorption condition (in all cases the transmittance was $T = 0$ due to the thick Au bottom layer).

Figure~\ref{f3} shows the thermal radiation spectra of multi-layer metasurfaces heated up to a 300$^{\circ}$C temperature. Due to the limited detection range of the HgCdTe (MCT) bolometer detector installed in the Fourier transform IR (FT-IR) spectrometer used for the measurements, only the spectral range from 1.5 to 10~$\mu$m could be probed. In this wavelength region the experimentally obtained emission and reflection spectra were in close accordance to the reciprocity expected from the thermodynamic equivalence between absorbance and emittance $E=A$. The peak wavelengths of reflectance matched well with emittance dips and \emph{vice versa}. Emissivity in the longer wavelength range becomes close to 95\% of that expected for black body radiation. In most cases the emittance of the second mode, which appeared around 2 to 4~$\mu$m became smaller than that of the first mode. In this region, there is no absorption by Si and SiO$_2$, however, energy might be lost to the coupling into propagation of plasmon or other lossy lateral modes supported by the planar waveguide formed between the two metal coatings (Fig.~\ref{f1}(a)).
\begin{figure}[tb]
\begin{center}
\includegraphics[width=12cm]{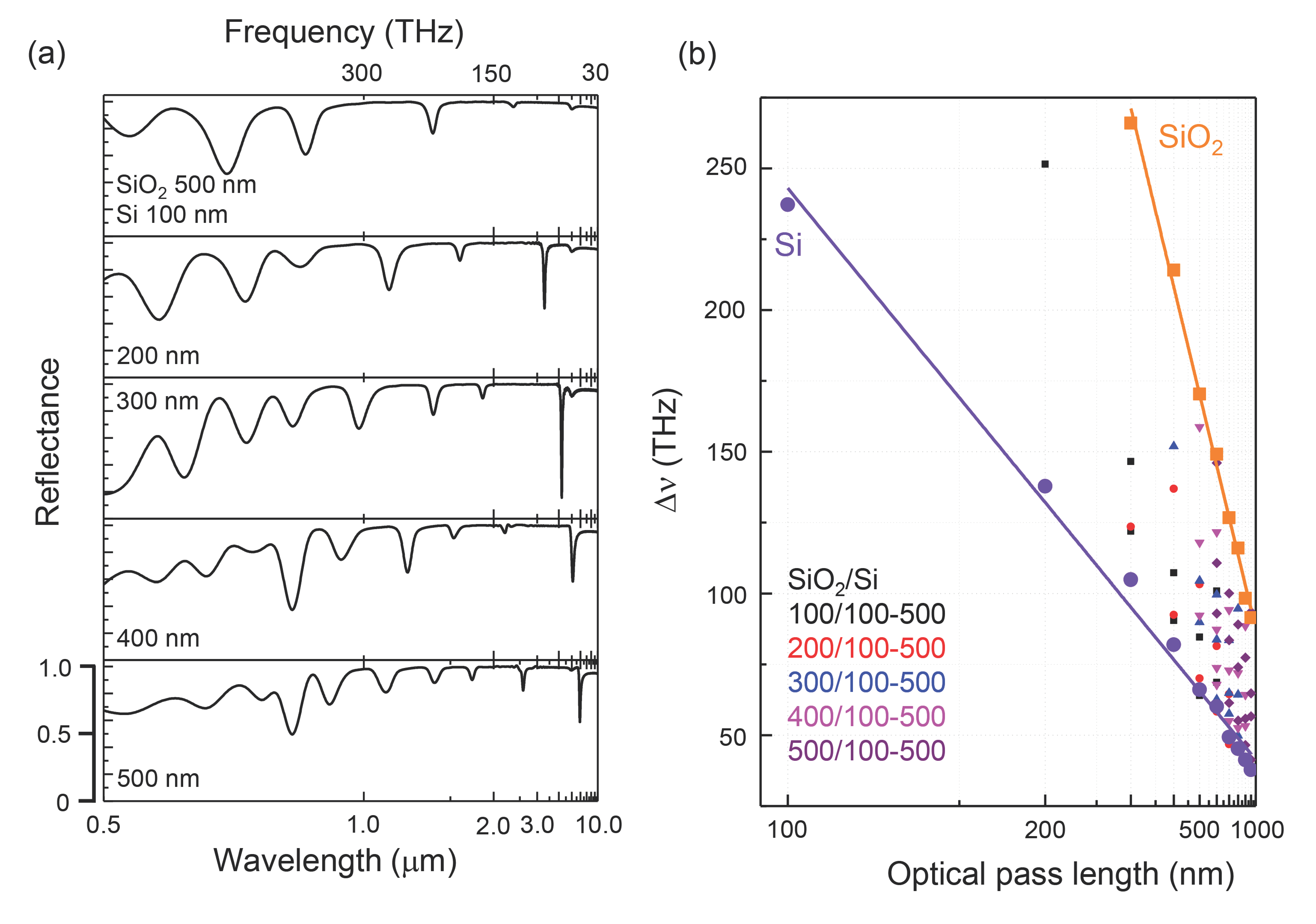}
\caption{FDTD simulations (Lumerical) of multilayered metasurfaces. (a) Reflection spectra of the SiO$_2$ 500~nm and Si layer with thickness spanning 100--500~nm. (b) The FSR of the SiO$_2$ and Si with range of thicknesses from 100 to 500~nm metasurfaces; the bounding lines represents pure Si and SiO$_2$ cases.} \label{f4}
\end{center}
\end{figure}

\subsection{Numerical simulations}

Finite difference time domain (FDTD) modeling is well suited to simulate optical properties of the nano-layered metasurfaces. Interestingly, for the fabricated multi-layer structures, the FDTD results were not conclusive at the expected quantitative level. The FDTD data are summarised in Fig.~\ref{f4}. In FDTD simulation, FSR became a complex number. Pure Si and SiO$_2$ generate relatively uniform and regularly spaced peaks over the simulated frequency range. All the FSRs should be located between the lines for the pure Si and SiO$_2$ (Fig.~\ref{f4}(b)). The complex FSR numbers hints that the multi-reflections had to occur at the boundaries of Au/Si, Si/Sio$_2$ and SiO$_2$/Au. In experiments however, the peaks were broader and there were less of them as obtained by FDTD (Fig.~\ref{f4}(a)).

We used incoherent and un-polarized light in experiment which is different from the FDTD simulations where coherent effects are captured. Moreover, the actual samples are expected to have some roughness on the surfaces and interfaces which otherwise were ideal mirror-like in simulations. Therefore the optical properties determined experimentally became broadened and difficult to completely reproduce by FDTD, which  used a coherent, polarized light with smooth surfaces. More detailed study of correspondence between experiment and FDTD simulations were reported in the previous study~\cite{18anm3557}. It is noteworthy, that FDTD codes for the slow light modes (large $n_g$) are specially developed in order to capture absorption and interference in photonic crystal for light trapping in solar cells at the edge of absorption band~\cite{Sajeev}. Hence, a standard commercial FDTD software is not optimised for the slow light absorption.    

\section{Conclusions and outlook}

A simple two-layer Si-SiO$_2$ structure placed between Au layers acts as an absorber and thermal emitter over the IR-MIR spectral window which is widely used for molecular fingerprinting in sensor applications. Close to perfect absorber performance was achieved at longer mid-IR wavelength range with emittance also approaching 95\%  of the black body radiation. The atmospheric transmission window with $T\approx 90\%$ between 3 and 4~$\mu$m wavelengths and slightly lower transmisivity at the 8-13~$\mu$m band can be utilised for radiative cooling using simple layered metasurfaces.    

\section*{Acknowledgements}
\small{This research was funded by Japan Society for the Promotion of Science (JSPS), Grants-in-Aid for Scientific Research, JSPS Bilateral Joint Research Projects between Japan and Lithuania JSPSBP120194203. SJ acknowledges partial support by the ARC Linkage  LP190100505 and JST CREST JPMJCR19I3 grants.
}
\bibliography{sample}






\end{document}